\documentclass[pss,fleqn]{w-art}
\usepackage{times}
\usepackage{w-thm}
\usepackage[]{graphicx}
\begin{document}
\DOIsuffix{theDOIsuffix}
\Volume{XX}
\Issue{1}
\Month{01}
\Year{2003}
\pagespan{3}{}
\Receiveddate{15 November 2003}
\Reviseddate{30 November 2003}
\Accepteddate{2 December 2003}
\Dateposted{3 December 2003}



\title[Removing spin blockade by  photon-assisted tunneling in double quantum dots.]
{Removing spin blockade by photon-assisted tunneling in double quantum dots.}


\author[R. S\'anchez]{Rafael S\'anchez\footnote{Corresponding
     author: e-mail: {\sf rafael.sanchez@icmm.csic.es}, Phone: +34\,913\,721\,420,
     Fax: +34\,913\,720\,623}\inst{1}}
\author[G. Platero]{Gloria Platero\inst{1}}
\author[R. Aguado]{Ram\'on Aguado\inst{1}}
\author[E. Cota]{Ernesto Cota\inst{2}}
\address[\inst{1}]{Instituto de Ciencia de Materiales de Madrid-CSIC, Cantoblanco, 28049, Spain}
\address[\inst{2}]{Centro de Ciencias de la Materia Condensada-UNAM, Ensenada B.C., M\'exico}
\begin{abstract}
Pauli exclusion principle can lead to a supression of transport through double quantum dots, even if both quantum dots are connected by a
resonant driving ac field. In this work, we study how this effect, known as spin blockade, can be removed by means of the influence of the
driving field on the contacts to the leads.
\end{abstract}
\maketitle                   





\section{Introduction}
One of the main properties of few electron quantum dots is the supression of charge current when the energy cost of introducing 
an extra electron in the system is large due to the charge repulsion inside the quantum dot (QD). In such a case, the system presents Coulomb
blockade, showing characteristic curves where current only flows for certain values of the chemical 
potentials\cite{Beenakker}. Between these peaks, 
current is allowed only through second order processes, which we do not treat here.

However, by introducing an external AC field, the electrons obtain (from the interaction with the field) enough energy to 
fulfill the energy requirements and tunnel through the system\cite{TienGordon}. In this way, a finite current may appear  
even in the zero bias configuration ({\it pumping} regime). 
These photon-assisted tunneling (PAT) processes through the contact barriers have been studied in AC driven single quantum
dots\cite{KouwenhovenPAT} but are usually 
neglected in the theoretical study of double quantum dot (DQD) resonant pumps\cite{StaffordWingreen}, where 
only interdot PAT is considered.

If one takes into account the spin of the electron, another interesting effect takes part in two-site systems like, for example, a DQD 
having one level each. 
If an electron is trapped in one of the quantum dots, transport is only available through the two-electron singlet state of that QD, when 
an electron with the opposite spin tunnels from the other QD and afterwards to the electrode. 
But, if the other 
QD is occupied by a trapped electron with the same spin, Pauli exclusion principle does not allow the formation of the doubly occupied state 
and, therefore, the current is blocked. This phenomenon is known as spin blockade (SB)\cite{Weinmann}. 

In this work, we study a concrete case in which spin blockade may appear in AC driven DQD spin pumps\cite{CotaNanot} and show that PAT 
processes through the contacts can be important, allowing the trapped spins to tunnel out of the system breaking the SB effect.

\section{Theoretical model}

Our system, consisting in two quantum dots weakly connected in series to two {\it unbiased} electron reservoirs by tunnel barriers, is described by the 
Hamiltonian $\hat{H}=\hat{H}_L+\hat{H}_R+\hat{H}_{L\Leftrightarrow R}+\hat{H}_{leads}+\hat{H}_T$. We consider one level in each QD containing 
up to two electrons: 
$\hat H_{j=\{L,R\}}=\sum_\sigma\varepsilon_{{j}\sigma}\hat{c}_{{j}\sigma}^\dagger\hat{c}_{j\sigma}+
U_{j}\hat{n}_{j\uparrow}\hat{n}_{j\downarrow}$, where $\hat{c}_{{j}\sigma}^\dagger$ is the creation operator of an electron with
spin $\sigma$ in dot $j$ and  energy $\varepsilon_{{j}\sigma}$ and $U_{j}$ is the charging energy
of each dot. The energies $\varepsilon_{{j}\sigma}$ include the Zeeman splitting, $\Delta_{j}$, due to the introduction of a 
magnetic field in order to break the spin degeneracy of the different levels, in such a way that we can consider the spin-up state 
as the ground state. Thus: $\varepsilon_{j\downarrow}=\varepsilon_{j\uparrow}+ \Delta_{j}$.
The reservoirs are described by the term: 
$\hat{H}_{leads}=\sum_{l\epsilon\{L,R\}k\sigma}\varepsilon_{lk}\hat{d}_{lk\sigma}^\dagger\hat{d}_{lk\sigma}$, where the operator 
$\hat{d}_{lk\sigma}^\dagger$ creates an electron with moment $k$ and spin $\sigma$ in lead $l$. Each QD is coupled to the 
other and to the leads through the terms:
$\hat{H}_{L\Leftrightarrow R}=-t_{LR}\sum_{\sigma}\hat{c}_{L\sigma}^\dagger\hat{c}_{R\sigma}+h.c.$ and 
$\hat H_T=\sum_{l\epsilon\{L,R\}k\sigma}(\gamma\hat{d}_{lk\sigma}^\dagger\hat{c}_{l\sigma}+h.c.)$. The constant 
that describes the tunneling through the contacts, $\gamma$, is asumed to be small and, for simplicity, similar for both QDs and will be treated as 
a perturbative parameter. 

We also introduce an external AC field acting on the energy levels of the quantum dots such that (considering $\hbar=e=1$):
$\varepsilon_{L(R)\sigma}\rightarrow\varepsilon_{L(R)\sigma}(t)=\varepsilon_{L(R)\sigma}\pm \frac{V_{AC}}{2}cos\omega t$, where 
$V_{AC}$ and $\omega$ are the amplitude and frequency of the field, respectively. The time dependent field will serve as a driver 
of electrons from one QD to the other, allowing the formation of doubly occupied states, which contribute to the current 
through the device. 

\subsection{Master equation}
We study the electron dynamics of the DQD system using the reduced density matrix operator, $\hat\rho=tr_R\hat\chi$, obtained by 
tracing all the reservoir states in the density operator of the whole system, $\hat\chi$. The Liouville equation, 
$\dot{\hat\rho}(t)=-i[\hat H(t),\hat\rho(t)]$ gives us the time evolution of the system. Assuming Markov and Born 
approximations\cite{Blum}, we derive the master equation for the density matrix elements\cite{pssa}, 
$\rho_{m'm}(t)=\langle m'|\hat{\rho}(t)|m\rangle$:
\begin{eqnarray}
\dot\rho(t)_{m'm}&=&-i\omega_{m'm}\rho_{m'm}(t)-i[\hat H_{L\Leftrightarrow R}'(t),\hat\rho(t)]_{m'm}
\\
&&+\left(\sum_{n\ne m'}\Gamma_{m'n}\rho_{nn}(t)-\sum_{n\ne m}\Gamma_{nm}\rho_{mm}(t)\right)\delta_{m'm}
-\Omega_{m'm}\rho_{m'm}(t)(1-\delta_{m'm}),\nonumber
\label{mastereq}
\end{eqnarray}
in the particle basis: 
$|1\rangle=|0,0\rangle$, $|2\rangle=|\uparrow,0\rangle$, $|3\rangle=|\downarrow,0\rangle$,
$|4\rangle=|0,\uparrow\rangle$, $|5\rangle=|0,\downarrow\rangle$, $|6\rangle=|\uparrow,\uparrow\rangle$,
$|7\rangle=|\downarrow,\downarrow\rangle$, $|8\rangle=|\uparrow,\downarrow\rangle$,
$|9\rangle=|\downarrow,\uparrow\rangle$, $|10\rangle=|\uparrow\downarrow,0\rangle$,
$|11\rangle=|0,\uparrow\downarrow\rangle$, $|12\rangle=|\uparrow\downarrow,\uparrow\rangle$,
$|13\rangle=|\uparrow\downarrow,\downarrow\rangle$, $|14\rangle=|\uparrow,\uparrow\downarrow\rangle$,
$|15\rangle=|\downarrow,\uparrow\downarrow\rangle$,
$|16\rangle=|\uparrow\downarrow,\uparrow\downarrow\rangle$.
Here, $\omega_{m'm}$ is the energy difference between the states $|m'\rangle$ and $|m\rangle$ of the isolated DQD and
$\Omega_{m'm}$ describes the decoherence of the DQD states due to the interaction with the reservoir. The time dependence 
of the energy levels has been transferred to the interdot coupling:
\begin{equation}
\langle m|\hat H'_{L\Leftrightarrow R}(t)|n\rangle=
\sum_{\nu=-\infty}^\infty J_\nu(\alpha)e^{i\nu\omega t}\langle
m|\hat H_{L\Leftrightarrow R}(t)|n\rangle 
\label{hoppingTrans},
\end{equation}
where $J_\nu(\alpha)$ is the $\nu$-th order Bessel function of the first kind, being
$\alpha=V_{AC}/\omega$ the dimensionless AC field intensity. 
The tunneling rates through the contacts, $\Gamma_{mn}$, are affected by the AC field: 
\begin{equation}
\Gamma_{mn}=\sum_{\nu=-\infty}^\infty J_\nu^2\left(\frac{\alpha}{2}\right) \xi_{mn}(\omega_{mn}+\nu\omega),
\label{patrates}
\end{equation}
where 
\begin{equation}
\xi_{mn}(\varepsilon)=\left\{ f(\varepsilon)\delta_{N_m,N_n+1}
+(1-f(-\varepsilon))\delta_{N_m,N_n-1}\right\}\Gamma
\label{rates}
\end{equation}
are the usual tunneling rates obtained {\it when PAT through the contacts is neglected}.  
$N_k=\sum_{j\sigma}\langle k|\hat n_{j\sigma}|k\rangle=\sum_j N_k^j$ is the number of electrons in the system in
state
$|k\rangle$, $f(\varepsilon)=1/(1+e^{(\varepsilon-\mu)\beta})$ is the Fermi distribution function, where $\beta=1/k_BT$ and $\mu$ is the chemical potential of the leads, and $\Gamma=2\pi|\gamma|^2$. 
These transition rates (Eq.(\ref{patrates})) are related to the decoherence through the 
relation:
$\Re\Omega_{m'm}=\frac{1}{2}(\sum_{k\ne
m'}\Gamma_{km'}+\sum_{k\ne m}\Gamma_{km})$.

We obtain the current that flows through the right contact with the relation 
$I_R=\sum_{m,m'}(\Gamma_{m'm}\rho_{mm}-\Gamma_{mm'}\rho_{m'm'})\delta_{N_m^R-1,N_{m'}^R}=\sum_{\sigma} I_{R,\sigma}$. The spin currents, $I_{R,\uparrow}$ and $I_{R,\downarrow}$, only account for the processes involving the tunneling of spins with {\it up} and {\it down} polarization, respectively. 

\section{Numerical results}

If both reservoirs have the same chemical potential, $\mu$, i.e., there is no bias voltage applied to the DQD, and 
$\mu<U_l+\varepsilon_l+\Delta_l$, the non-driven system will be in a stable state if it contains one electron in each dot. Additionally, 
if $\mu>U_R+\varepsilon_R$, the spin down electron will be the only one 
able to tunnel out to the right lead from the doubly occupied singlet state and the spin up will be trapped in the right QD. Thus, the breaking of the spin degeneracy by the introduction of a magnetic field leads to an asymmetry in the transport properties of the system, so the spin components of the current, $I_{R,\uparrow}$ and $I_{R,\downarrow}$, will behave differently.

\begin{figure}[htb]
\includegraphics[width=3.7in,clip]{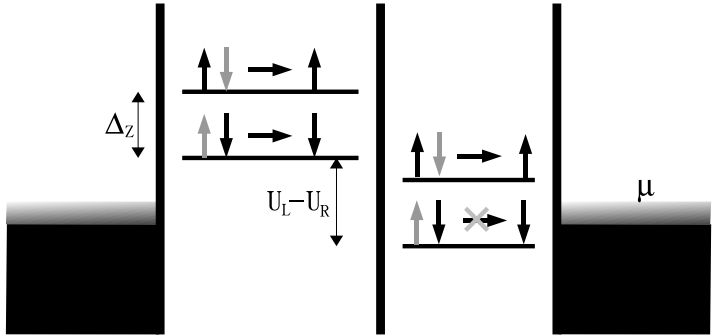}
\caption{\label{esquema} 
{\small Schematic diagram of the non-driven device showing the chemical potentials associated to the transitions involving the extraction through the contact barriers of one electron from the doubly occupied state in each QD. Since $\mu>U_R+\varepsilon_R$, the transitions extracting an electron with spin-up polarization from states with two electrons in the right dot through the right contact are energetically unavailable, unless they are mediated by the absorption of photons. In our configuration, the Zeeman splitting and the energy of the levels are the same in both dots, that is: $\Delta_L=\Delta_R=\Delta_z$ and $\varepsilon_L=\varepsilon_R$.
}}
\end{figure}

Introducing an AC field in resonance with the states $|\downarrow,\uparrow\rangle$ 
and $|0,\uparrow\downarrow\rangle$ (and, since $\Delta_L=\Delta_R$, also with $|0,\uparrow\downarrow\rangle$ and $|\uparrow,\downarrow\rangle$), 
the spin down electron will be delocalized between both quatum dots and there will be a finite 
probability for it to leave the DQD to the right lead. If PAT processes 
through the contacts are not considered, one should expect a net spin down current 
through the system (through the {\it pumping cycle}: 
$|\downarrow,\uparrow\rangle\Leftrightarrow|0,\uparrow\downarrow\rangle\rightarrow\left\{|0,\uparrow\rangle or
|\downarrow,\uparrow\downarrow\rangle\right\}\rightarrow|\downarrow,\uparrow\rangle$) but, since the empty left QD can be also filled with a spin up electron, the system asymptotically evolves to the state $|\uparrow,\uparrow\rangle$
(i.e., $\rho_{6,6}(t\rightarrow\infty)=1$ and $\rho_{i,j}(t\rightarrow\infty)=0$, otherwise) 
that leads to SB\cite{CotaNanot}.

\begin{figure}[htb]
\includegraphics[angle=270,width=3.7in,clip]{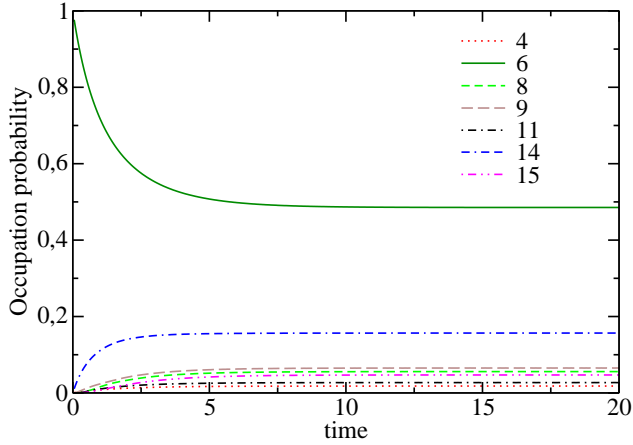}
\caption{\label{dens} 
{\small Time evolution (normalized to $\tau=2\pi/\Omega$, being $\Omega=2J_1(\alpha)t_{LR}$ the Rabi frequency of the delocalization processes) of the diagonal density matrix elements $\rho_{4,4}$, $\rho_{6,6}$, $\rho_{8,8}$, $\rho_{9,9}$, 
$\rho_{11,11}$, $\rho_{14,14}$ and $\rho_{15,15}$, 
which describe the occupation probability of the states 
that contribute to the current. We consider the initial condition: $\rho_{6,6}(t=0)=1$. In the case where PAT through the contacts 
is not considered, we would obtain $\rho_{6,6}(t)=1$, for all times\cite{CotaNanot}.
We do not show the occupation probabilities of other states that are not directly involved in the pumping processes.
Parameters (in meV): $t_{LR}=0.005$, $\Gamma=0.001$, $U_L=1.6$, $U_R=1.3$,
$\Delta_L=\Delta_R=0.2$ (corresponding to a magnetic field $B\approx8T$), $\varepsilon_L=\varepsilon_R=0.5$, $\mu=1.9$, 
$\omega=\omega_{11,8}=V_{AC}/2$. 
}}
\end{figure}

\begin{figure}[htb]
\includegraphics[angle=270,width=3.7in,clip]{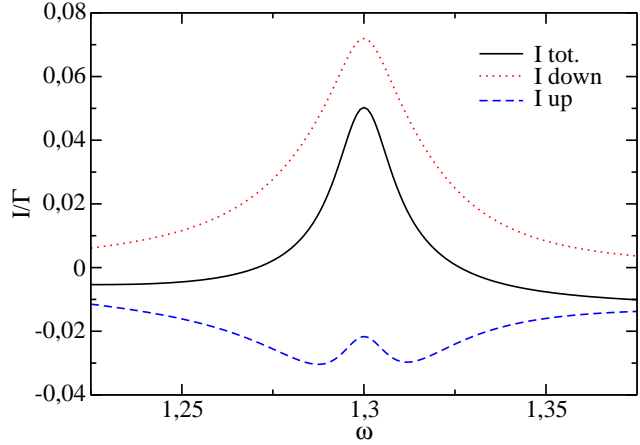}
\caption{\label{Ivsw} 
{\small Pumped current (normalized to the tunneling probability, $\Gamma$) as a function of the frequency of the AC field (in meV). 
Considering the {\it blocking state}, $|\uparrow,\uparrow\rangle$, if the electron in the left QD absorbs a photon and tunnels to the left lead, net spin down current to the right lead is created due to the sequence: 
$|\uparrow,\uparrow\rangle\rightarrow|0,\uparrow\rangle\rightarrow|\downarrow,\uparrow\rangle\Leftrightarrow|0,\uparrow\downarrow\rangle
\rightarrow\{|0,\uparrow\rangle or |\uparrow,\uparrow\downarrow\rangle\}\rightarrow|\uparrow,\uparrow\rangle$, while spin up current 
flows in the opposite direction (from right to left) by the sequence:
$|\uparrow,\uparrow\rangle\rightarrow|0,\uparrow\rangle\rightarrow|\downarrow,\uparrow\rangle\Leftrightarrow|0,\uparrow\downarrow\rangle
\Leftrightarrow|\uparrow,\downarrow\rangle\rightarrow|\uparrow,\uparrow\downarrow\rangle\rightarrow|\uparrow,\uparrow\rangle$ (in this 
cycle, spin down current through the right contact is also produced). 
The double arrow ($\Leftrightarrow$) represents the resonant delocalization processes inside the DQD at $\omega\approx U_R=1.3$. 
On the other hand, if the spin up in the right QD is extracted from $|\uparrow,\uparrow\rangle$, there is a positive contribution to spin up current through the sequence:
$|\uparrow,\uparrow\rangle\rightarrow|\uparrow,0\rangle\rightarrow|\uparrow,\downarrow\rangle\Leftrightarrow|0,\uparrow\downarrow\rangle\rightarrow\{|\uparrow,\uparrow\downarrow\rangle or |0,\uparrow\rangle\}\rightarrow|\uparrow,\uparrow\rangle$. The contribution of this sequence is smaller (since it compites with the sequence $|\uparrow,\uparrow\rangle\rightarrow|\uparrow,0\rangle\rightarrow|\uparrow,\downarrow\rangle\rightarrow|\uparrow,\downarrow\rangle\rightarrow|\uparrow,\uparrow\downarrow\rangle\rightarrow|\uparrow,\uparrow\rangle$ that recovers rapidly the state $|\uparrow,\uparrow\rangle$ without contributing to the current) and is only apreciable at high enough field intensities. However, it may be the responsible of the supression of negative spin up current near resonance, giving a small bump. Note also that, since so many states are contributing to the dynamics of the system, the behaviour of the resonance peaks differs from the typical Lorentzian shape.
The parameters are the same as in Fig. \ref{dens}.
}}
\end{figure}

However, the rates (\ref{patrates}) allow the "trapped" spins in the DQD to absorb a certain number of photons and 
tunnel out to the leads giving a finite ocuppation probability (through the sequences: 
$|\uparrow,\uparrow\rangle\rightarrow|0,\uparrow\rangle\rightarrow|\downarrow,\uparrow\rangle$ or $|\uparrow,\uparrow\rangle\rightarrow|\uparrow,0\rangle\rightarrow|\uparrow,\downarrow\rangle$) 
to the states $|\downarrow,\uparrow\rangle$ and $|\uparrow,\downarrow\rangle$ which are in resonance with the state $|0,\uparrow\downarrow\rangle$, that contributes to the pumping of a spin down electron to the right lead (Fig. \ref{dens}).
Then, PAT through the contacts creates a finite current through the system (Fig. \ref{Ivsw}), removing the SB.

\section{Conclusions}
We have shown that the interaction with the driving field applied on a DQD system affects not only the interdot tunneling 
but also the tunneling through the contact barriers. In concrete, PAT through the contact barriers affects the ocupation 
of the states, giving a finite probability to states that are energetically unstable but open new channels to the electronic 
transport through the system. Therefore, they should be taken into account when studying properties of driven quantum dot devices.

\begin{acknowledgement}
Work supported by Programa de Cooperaci\'on Bilateral CSIC-CONACYT, by Grant No. DGAPA-UNAM
114403-3, by the EU Grant No. HPRN-CT-2000-00144 and by the Ministerio de Educaci\'on y Ciencia of Spain 
through Grant No. MAT2005-00644. RS was supported by CSIC-Programa I3P, cofinanced by Fondo Social Europeo.
\end{acknowledgement}

\end{document}